\documentclass[twocolumn]{aastex631}

\usepackage{slashbox,pict2e}
\usepackage{diagbox}

\usepackage{graphicx}
\usepackage{amssymb}
\usepackage{amsmath}
\usepackage{graphicx}
\usepackage{amstext}
\usepackage{leftidx}
\usepackage{esint}
\usepackage[utf8]{inputenc}
\usepackage{color}
\usepackage{hyperref}
\usepackage{multido}

\newcommand{\CC}{\Lambda}

\newcommand{\rv}{\rho_{\rm vac}}

\newcommand{\rvo}{\rho^0_{\rm vac}}

\newcommand{\rmr}{\rho_m}

\newcommand{\rX}{\rho_X}

\newcommand{\wX}{\omega_X}
\newcommand{\wY}{\omega_Y}

\newcommand{\CCS}{\CC_s{\rm CDM}}
\newcommand{\wXCDM}{$w${\rm XCDM}\,}

\newcommand{\nueff}{\nu_{\rm eff}}

\newcommand{\mpl}{m_{\rm Pl}}

\newcommand{\be}{\begin{equation}}
\newcommand{\ee}{\end{equation}}

\begin{document}

\title{Phantom matter: a challenging solution to the cosmological tensions}

\author[0000-0002-2922-2622]{Adri\`a G\'omez Valent}
\affiliation{Departament de F\'isica Qu\`antica i Astrof\'isica and Institute of Cosmos Sciences,\\ Universitat de Barcelona,
Av. Diagonal 647, E-08028 Barcelona, Catalonia, Spain}

\author[0000-0002-5295-8275]{Joan Sol\`a Peracaula}
\affiliation{Departament de F\'isica Qu\`antica i Astrof\'isica and Institute of Cosmos Sciences,\\ Universitat de Barcelona,
Av. Diagonal 647, E-08028 Barcelona, Catalonia, Spain}


\begin{abstract}
The idea of composite dark energy (DE)  is quite natural since on general grounds we expect that the  vacuum energy density (associated with the cosmological term  $\Lambda$) may appear in combination with other effective forms of DE, denoted $X$.  Here we deal with  model $w$XCDM, a simplified version of the old  $\Lambda$XCDM model (\cite{Grande:2006nn}),  and exploit the possibility that  $X$ behaves as   `phantom matter' (PM), which appears in stringy versions of the running vacuum model (RVM). Unlike phantom DE,  the PM fluid satisfies the strong energy condition like usual matter, hence bringing to bear positive pressure at the expense of negative energy. Bubbles of PM may appear in the manner of a transitory  `phantom vacuum' tunneled into the late universe before it  heads towards a new de Sitter era, thereby offering a crop field for the growing of structures  earlier than expected. Using  SNIa, cosmic chronometers, transversal BAO (BAO 2D), LSS data and the full CMB likelihood from Planck 2018, we find that  the $H_0$ and growth tensions virtually disappear, provided that BAO 2D are the only source of BAO data used in the fit. In contrast, our preliminary analysis using exclusively anisotropic BAO (BAO 3D) indicates that the ability to ease the $H_0$ tension is significantly reduced as compared to the scenario with BAO 2D, despite the fact that the overall fit to the cosmological data is still better than in the $\CC$CDM.  Finally, our approach with BAO 2D favors quintessence-like behavior of the DE below $z\simeq 1.5$ at  $\gtrsim 3\sigma$ CL, which is compatible with the recent DESI measurements.
\end{abstract}




\section{Introduction} \label{sec: intro}

The standard (or concordance) cosmological model, aka  $\CC$CDM, has been a rather successful  paradigm for the description of the universe for more than three decades\,\citep{peebles:1993}, especially since the late nineties\,\citep{Turner:2022gvw}. Despite it constitutes the main theoretical pillar at our hands for the description of the universe's evolution within  the General Relativity (GR) context, the $\CC$CDM is being pestered by a number of  glitches and hitches which prove to be more and more difficult to iron out.  While the model is presumably right in the main (at least at a pure phenomenological level), an increasing number of worrisome inconsistencies (or `` tensions'') have been perturbing its reputation and its future prospects. For a long time the role played by a rigid cosmological constant (CC) $\CC$  in the standard $\CC$CDM model has been rather successful since it  has provided a fairly reasonable description of the overall cosmological data\,\citep{Planck:2015fie,Planck:2018vyg}. But the use of the cosmological term has never been fully clarified at the formal theoretical level since it is usually associated to the so-called ``cosmological constant problem'', a serious conundrum which has been amply discussed in the literature, see e.g. \citep{Weinberg:1988cp,Peebles:2002gy,Padmanabhan:2002ji,SolaPeracaula:2022hpd,Sola:2013gha}  and references therein. Inasmuch as the vacuum energy density (VED) is related to $\CC$ through $\rv=\CC/(8\pi G_N)$ ($G_N$ being Newton's gravitational coupling), the conceptual fate of $\CC$ is tied to our ultimate understanding of the VED on fundamental physical terms.  Fortunately, recent theoretical developments on the VED in quantum field theory (QFT) are bringing new light for a potential alleviation of these theoretical difficulties. In fact, the possibility that the quantum vacuum (and hence that $\CC$ itself) is actually dynamical (i.e. evolving with the cosmological expansion) rather than being stuck at a rigid value, has lately been  substantiated in the context of QFT in curved spacetime \citep{SolaPeracaula:2022hpd,Sola:2013gha} as well as in the framework of low-energy effective string theory \citep{Mavromatos:2020kzj,Mavromatos:2021urx}  and, very recently, in  lattice quantum gravity\,\citep{Dai:2024vjc}. For a long time, the dynamics of the VED has been popularly addressed on phenomenological grounds  using ad hoc scalar fields (quintessence and the like) which supplant the role of the $\CC$-term through the current value of some suitable effective potential \citep{Peebles:2002gy,Padmanabhan:2002ji}, see e.g. \citep{Avsajanishvili:2023jcl} for a recent review.  Currently, the first data release of the Dark
Energy Spectroscopic Instrument (DESI) suggests tantalizing evidence that the dark energy (DE) might be dynamical using some common parameterizations of the DE \citep{DESI:2024mwx}. More detailed analyses will be needed before getting a final confirmation, of course,  but it is a fact that some anticipatory (and fairly robust) hints of dynamical DE in the literature already pointed out this possibility a few years ago  from different perspectives and using a significant amount of cosmological data. The level of evidence put forward by these  anticipatory studies was substantial and ranged $3-4\sigma$.  Some of these analysis are well-known\,\citep{Sola:2015wwa,Sola:2016jky,Sola:2017znb,SolaPeracaula:2016qlq,SolaPeracaula:2017esw,Gomez-Valent:2018nib} and involved the running vacuum model (RVM)\footnote{For a review, see e.g. \citep{SolaPeracaula:2022hpd,Sola:2013gha,Sola:2015rra} and references therein.}. Subsequent analysis around the same time supported also this possibility using different methods and parameterizations\,\citep{Zhao:2017cud,SolaPeracaula:2018wwm}.

Dynamical DE could also impinge positively on the resolution of the cosmological tensions. Let us recall that these are mainly concerned with the measurement of the current Hubble parameter $H_0\equiv 100 h$ km/s/Mpc and the growth of large scale structures (LSS). The latter is usually monitored with $S_8$ or $\sigma_8$, or even better using a parameter $\sigma_{12}$ defining the amplitude of the
matter power spectrum at fixed spheres of radius $12$ Mpc rather than $8h^{-1}$ Mpc, thus avoiding artificial dependence on the value of $h$\,\citep{Sanchez:2020vvb,eBOSS:2021poy,Semenaite:2022unt}\footnote{See \citep{DiValentino:2020zio,DiValentino:2020vvd} for summarized explanations about each of these tensions, and \citep{Perivolaropoulos:2021jda,Abdalla:2022yfr} for comprehensive reviews.}.  The first kind of tension involves a  serious disagreement  between the  CMB observations, using fiducial $\CC$CDM cosmology, and the local direct (distance ladder) measurements of the Hubble parameter today. It is arguably the most puzzling open question  and it leads to a severe inconsistency of $\sim 5 \sigma$ c.l. between the mentioned observables. The second kind of tension is related with the exceeding rate of large scale structure formation in the late universe predicted by the $\CC$CDM as compared to measurements, although the discordance here is moderate,  at a confidence level of $\sim 2-3\sigma$.   More recently,  data from the James Webb Space Telescope (JWST)\,\citep{Gardner:2006ky,Labbe:2022ahb} have revealed the existence of an unexpectedly large population of extremely massive galaxies at large redshifts $z\gtrsim 5-10$, a fact which is also strongly at odds with the prospects of the  concordance model.

There is a wide panoply of  strategies in the literature trying to mitigate some of the above tensions, although at present we may be still far away from a satisfactory resolution of the situation. We mention only a few. It has been argued that within the class of models where the DE is dealt with as a cosmic fluid with equation of state (EoS) $w(z)$, solving the $H_0$ tension demands the phantom condition $w(z)<-1$ at some $z$, whereas solving both the $H_0$ and $\sigma_8$ tensions requires $w(z)$ to cross the phantom divide and/or other sorts of exotic transitions, see e.g. \citep{Heisenberg:2022gqk,Marra:2021fvf,Alestas:2020zol,Perivolaropoulos:2021bds,Alestas:2021luu,Perivolaropoulos:2022khd,Gomez-Valent:2023uof}. The possibility of a sign flip of the $\CC$ term has been entertained in recent times,  e.g. in \citep{Calderon:2020hoc}. Let us also mention the model analyzed in \citep{Akarsu:2021fol,Akarsu:2023mfb} (based on the framework of \cite{Akarsu:2019hmw}), in which one considers a sudden transition from anti-de Sitter  (AdS), hence $\CC<0$, into de Sitter (dS)  regime ($\CC>0$) occurring near our time. Despite it being essentially ad hoc, the model yields a rather good fit to the data when transversal/angular BAO (BAO 2D for short) is employed in the fitting analysis, see also \citep{Gomez-Valent:2023uof}. Actually, a more general framework already existed since long ago, namely the  $\CC$XCDM model\,\citep{Grande:2006nn,Grande:2006qi,Grande:2008re}, where the running vacuum and an extra $X$ component can exchange energy. In such a context, one can have $\CC<0$ or $\CC>0$.

In this work, we further exploit the virtues of the $\CC$XCDM, which is an enhanced family of RVM's\,\citep{SolaPeracaula:2022hpd}. The linchpin of the RVM framework  is the dynamical nature of the VED framed in the fundamental context of QFT in curved spacetime or in low-energy string theory. Appropriate renormalization in this context gets rid of the quartic mass terms $\sim m^4$ responsible for the fine tuning troubles, and as a result the VED becomes a smooth function of even powers of the Hubble rate\,\citep{Moreno-Pulido:2020anb,Moreno-Pulido:2022phq,Moreno-Pulido:2022upl,Moreno-Pulido:2023ryo}.  This leads to a fairly good description of the cosmological data and helps easing the tensions \citep{SolaPeracaula:2021gxi,SolaPeracaula:2023swx}. Worth noticing is also the phenomenological performance of the stringy version of the RVM \citep{Gomez-Valent:2023hov}. This said, we cannot exclude that the DE may be a composite fluid, a possibility which could further help in the task of alleviating these tensions.  For this reason a good candidate is the aforementioned `$\Lambda$XCDM model'\,\citep{Grande:2006nn,Grande:2006qi,Grande:2008re}, a RVM-born model which was initially motivated as a possible cure for the cosmic coincidence problem. In it, we have two DE components: one is the running VED, $\rv=\rv(H)$, and the other is the mentioned $X$.  As a matter of fact, the entity $X$ need not be a fundamental field, as emphasized in \citep{Grande:2006nn}: it may just be due to particular terms in the effective action which mimic a dynamical DE component.  In this sense, $X$ can have phantom-like behavior without causing uproar at the theoretical level. Not only so, $X$ may even display ``phantom matter'' (PM) behavior \citep{Grande:2006nn}, namely an intriguing form of DE which, in stark contrast to the usual phantom DE, is characterized by a positive pressure ($p_X>0$) at the expense of a negative energy density ($\rX<0$). It is remarkable that there exists specific theoretical scenarios in the current literature which support the $\CC$XCDM structure with a PM-like component $X$. An example appears in the stringy RVM context \citep{Mavromatos:2020kzj,Mavromatos:2021urx}.  No less remarkable is the fact that the phenomenological performance attained by such a peculiar composite DE scenario may surpass by far that of the standard $\CC$CDM under appropriate conditions, as we shall shortly clarify.  However, for the sake of simplicity, in this paper we shall address a reduced version of the $\CC$XCDM, which we call the $w$XCDM model (not to be confused with the conventional $w$CDM parameterization), in which the additional $X$ component of the $\CC$XCDM is kept intact, but on the other hand we mimic the running $\CC$  with another dynamical component which we call $Y$.  While the analysis within the full $\CC$XCDM  will be presented elsewhere, we shall nonetheless demonstrate here that even the simplified  $w$XCDM model can be very efficient in dealing with the cosmological tensions. However, for this to be so  the following two conditions must be met in our analysis: i) the $X$ component must have the ability to behave as phantom matter; and ii)  BAO 2D must be the only source of BAO data used in our global fit.  The precise reasons for considering only this type of BAO data in the current work will be explained below, together with a preliminary discussion of the results obtained with the BAO 3D variant. The upshot is that by accepting these reasons for using BAO 2D only,  and also taking advantage of the mentioned  (stringy) theoretical framework to accommodate phantom matter, the possibility to quell the cosmological tensions on fundamental grounds is viable and can be extremely effective.


\section{Composite dark energy}\label{sec:CompositeDE}
We next mention three related types of composite models of the DE: i) $\CC$XCDM,  ii) \wXCDM and iii) $\CCS$. As noted,  model i) exists in different versions since long ago\,\citep{Grande:2006nn,Grande:2006qi,Grande:2008re};  model ii) will be analyzed here for the first time, it  constitutes a simplified version of model i) and embodies the PM feature. Finally, model iii) was recently analyzed in\,\citep{Akarsu:2023mfb}.

$\bullet$ i) $\CC$XCDM model. Its definition and comprehensive discussion, including the background cosmological solution, are provided in utmost detail in\,\citep{Grande:2006nn}. In addition, in \citep{Grande:2008re} the corresponding cosmic perturbations equations are fully accounted for. Here we just present a very short qualitative description. Within the $\CC$XCDM,  the cosmic fluid contains the usual matter  energy density $\rmr$  and a composite DE sector made out of  two components, one is the running vacuum energy density $\rv$ and the other is called $X$, with energy density $\rX$. The VED here is treated within the QFT framework of the RVM\,\citep{Moreno-Pulido:2020anb,Moreno-Pulido:2022phq,Moreno-Pulido:2022upl,Moreno-Pulido:2023ryo}, in which $\rv= \rv(H)$ evolves with the expansion rate.  In the current universe, such an evolution reads\,\citep{SolaPeracaula:2022hpd}:
\begin{equation}\label{eq:RVM}
\rv(H) = \rvo+\frac{3\nueff}{8\pi}\,(H^2-H_0^2)\,\mpl^2\,,
\end{equation}
with $\rvo$ the current VED value,  $\mpl$ the Planck mass  and  $|\nueff|\ll1$ a small  parameter which is formally computable in QFT, see the aforementioned papers. For $\nueff>0$ the VED decreases with the expansion and hence the RVM mimics quintessence, whereas for $\nueff<0$ the VED increases with the expansion and the RVM behaves effectively as phantom DE.
The measured cosmological term is $\CC=8\pi G\rv$, and hence $\CC$ also runs  with $H$ since $\rv=\rv(H)$.  Such a running feature of the VED and $\CC$ occurs in the $\CC$XCDM too,  and in exactly the same form \eqref{eq:RVM}, but here we have also the dynamics of $X$ and hence the cosmological solution of the model in terms of the redshift variable is considerably more complicated\,\citep{Grande:2006nn}.
Besides, as recently demonstrated in the QFT context, $\rv$ has an EoS which departs from $-1$ \citep{Moreno-Pulido:2022upl}, so the ``modern version of the $\CC$XCDM model'' has actually two nontrivial EoS parameters, one for the running VED and the other for $X$, which we call $\wX$. These are basic ingredients in our analysis,  and for this reason the $\CC$XCDM fleshes out the theoretical basis inspiring the current work. However, the actual scenario which we will analyze here is model $w$XCDM (see below). It is simpler than the $\CC$XCDM but it has the same genetic ingredients. We will use it to emulate the basic properties of the latter. The analysis of the cosmic tensions within the full $\CC$XCDM framework is more demanding and will be presented in a separate work \citep{LXCDM2024}.

$\bullet$  ii) \wXCDM model.  In order to illustrate the possibilities of our composite DE scenario in connection with the cosmological tensions, herein we will use the \wXCDM model, a reduced (skeleton) version of the $\CC$XCDM.  The two models share the $X$ component, but \wXCDM  mimics the running vacuum feature of $\CC$XCDM through another dynamical component $Y$ (replacing $\CC$) and whose EoS,  $w_Y$, can be different from $-1$. This means that $Y$ does not act as a rigid CC, which is fair enough since, as mentioned before,  in the RVM context the VED is dynamical and its EoS (hence that of $\CC$) actually  departs from $-1$ owing to quantum effects\,\citep{Moreno-Pulido:2022upl}. Thus, in the \wXCDM,  we have two DE components with potentially different EoS behaviors. Furthermore, as we will see in section \ref{sec:Discussion}, in the best-fit model the $X$ component behaves phantom-like ($\wX\lesssim -1$)  and the $Y$ component behaves quintessence-like ($\wY\gtrsim -1$). An important point is that $X$ and $Y$ do not act simultaneously: $X$ acts first in the cosmic expansion, and $Y$ acts subsequently. To be precise, $X$ enters only above a transition redshift $z>z_{t}$ (fitted from the data) whilst $Y$ enters below that redshift until the current time.
{No less crucial is the fact that the phantom-like component $X$ actually behaves as PM, therefore with negative energy density ($\Omega_X=\rX/\rho_c<0$) and positive pressure ($p_X>0$), whereas for the $Y$ component we have $\Omega_Y>0$, $p_Y<0$. Notice that the characteristic free parameters of model $w$XCDM are just $(z_{t}, w_X, w_Y)$. {Indeed, the density parameters for the DE components are not free since e.g. the  value of $\Omega_Y^0\equiv\Omega_Y(z=0)$ depends on the fitting values of $H_0, \omega_b, \omega_{\rm dm}$ (cf. Table \ref{tab:table_fits}).
In addition, the respective values of $\Omega_X$ and of $\Omega_Y$  immediately above and below $z_{t}$ are assumed to be equal in absolute value.  This means that $|\Omega_X(z)|=\Omega_Y(z)$ at $z=z_{t}$. This kind of assumption aims at reducing the number of parameters and is entirely similar to the one made in the $\CCS$ model, see point iii) below.}

Last, but not least, we note that the existence of these two different  phases of the DE separated by the redshift $z_{t}$ is not just an ad hoc assumption since it is found in theoretical contexts such as the stringy RVM approach\,\citep{Mavromatos:2021urx}, which points to the existence of transitory domains or bubbles of PM whenever the universe is approaching a de Sitter epoch (see next sections for more elaboration).

$\bullet$ iii) $\CCS$ model.  This is the model recently analyzed with considerable phenomenological success in \citep{Akarsu:2023mfb}.   Strictly speaking, it is not a composite model since only $\CC$ is involved, although it enters with two signs and in this sense it is a {composition of two phases of $\CC$ separated also by a transition redshift $z_t$, which is the characteristic parameter of this model.  It is assumed that at that point there is a sudden AdS-dS transition from $-\CC<0$ in the upper redshift range to $+\CC>0$ in the lower range, i.e. an abrupt change of sign of $\CC$ at $z=z_t$, but keeping the same absolute value. We refer the reader to the quoted reference for more details.

For convenience,  we shall simultaneously provide  the fitting results for models ii) and iii) under the very same data sources. Together with the standard $\CC$CDM model, the simultaneous analysis of $\CCS$ will be useful as a benchmark for rating the performance of the phantom matter approach $w$XCDM to the cosmological tensions. 

\section{Phantom matter and the energy conditions of the cosmic fluids}\label{sec:PMandPV}
Even though  phantom matter was first proposed phenomenologically in the old composite cosmological model $\CC$XCDM\,\citep{Grande:2006nn,Grande:2006qi,Grande:2008re},  more recently it has been put forward on more formal grounds in \citep{Mavromatos:2021urx}, specifically within the context of the stringy RVM approach, see \citep{Mavromatos:2020kzj} for a review.  Obviously, we need not provide technical details here and we refer the reader to the aforesaid references. However, we can at least describe the conceptual design of this picture.  In a nutshell, it is the following: in a framework of a string-inspired cosmology with primordial gravitational waves and gravitational anomalies, which lead to dynamical inflation of RVM type without the need for ad hoc inflaton fields\,\citep{Sola:2015rra,Mavromatos:2020kzj}, a crucial role  is played by the fundamental axion field existing in the gravitational multiplet of string theory, viz. the Kalb-Ramond (KR) axion field $b(x)$, which couples to the gravitational Chern-Simons (gChS) term: $\sim  b(x) \, R_{\mu\nu\rho\sigma}\, \widetilde R^{\mu\nu\rho\sigma}\equiv  b(x) R\,\tilde R$, where $\tilde R$  denotes the dual of the Riemann tensor. The KR axion,  when combined with the gChS contribution,  obey together a peculiar equation of state of vacuum type $p =-\rho$ but with negative energy density $\rho< 0$ (hence with positive pressure $p>0$), which defines the `phantom vacuum' \citep{Mavromatos:2021urx}. This form of vacuum, however, is merely transitory until the  gChS condensates $\langle R\,\tilde R\rangle$ become activated through the condensation of primordial gravitational waves (GW), thereby  making possible a positive-definite vacuum state energy with effective cosmological `constant' $\Lambda(H)\sim \langle b R\tilde{R}\rangle$.  At this juncture,  the total pressure and density already involve the combined effect from all the contributions, what makes possible {a normal vacuum state with  $\rho_{\rm total}>0$ and hence with $p_{\rm total}=-\rho_{\rm total} < 0$, i.e. a standard de Sitter phase with positive energy density and negative pressure}. The latter nevertheless may still be subdued to corrections owing to quantum effects. Interestingly enough, in such a framework the
KR axion can be a candidate to Dark Matter, what provides an additional motivation for this cosmological picture since it could embrace the entire cosmic history.

Thus, a phantom-matter dominated era can be present in the early cosmic evolution near a de Sitter phase, but we should emphasize that it could also reflourish in the late universe. As shown in \,\citep{Basilakos:2019mpe,Basilakos:2019acj,Basilakos:2020qmu}, owing to the generation of chiral matter, which will dominate the universe at the exit from inflation, at large scales the universe can recover its FLRW background profile. This is because the chiral matter fields  insure the cancellation of the gravitational anomalies during the FLRW regime, where they are indeed not observed as otherwise it would imply a glaring violation of general covariance. However, as soon as we are in the process of exiting the FLRW regime towards a new de Sitter era, we cannot exclude that in successive stages of the cosmic evolution the universe can be affected by the presence of  lurking PM. The reason is that the chiral matter fields will get more and more  diluted with the cosmic expansion and,  then,  owing to the incomplete cancellation of  the gravitational  anomalies, these may eventually re-surface  near  the  current quasi-de Sitter era.  It is therefore not inconceivable to think of the existence of phantom matter bubbles or domains tunneling now and then  into our universe at relatively close redshift ranges before the ultimate de Sitter phase takes over.  These bubbles of PM, being endowed with positive pressure $p>0$,  could  obviously foster a larger rate of the structure formation, a fact which might explain the overproduction of large scale structures at unexpected places and times deep in our past.  As we shall next show, this ideology provides an excellent framework for a fit to the overall cosmological data which proves to be (far) better than that of the standard $\CC$CDM model, provided that we exclusively focus on BAO 2D data in our fit.   We refer once more the reader  to \citep{Mavromatos:2021urx} and the review \citep{Mavromatos:2020kzj}  for the details underlying the PM scenario supporting our proposal for solving the tensions. See also \citep{Mavromatos:2021sew} and the recent developments \citep{Dorlis:2024yqw, Mavromatos:2024pho}.  Here we have limited ourselves to a very succinct exposition of the theoretical background and in the rest of this work we focus on exploring the phenomenological implications.


\begin{figure}[!t]
\begin{center}
\includegraphics[scale=0.067]{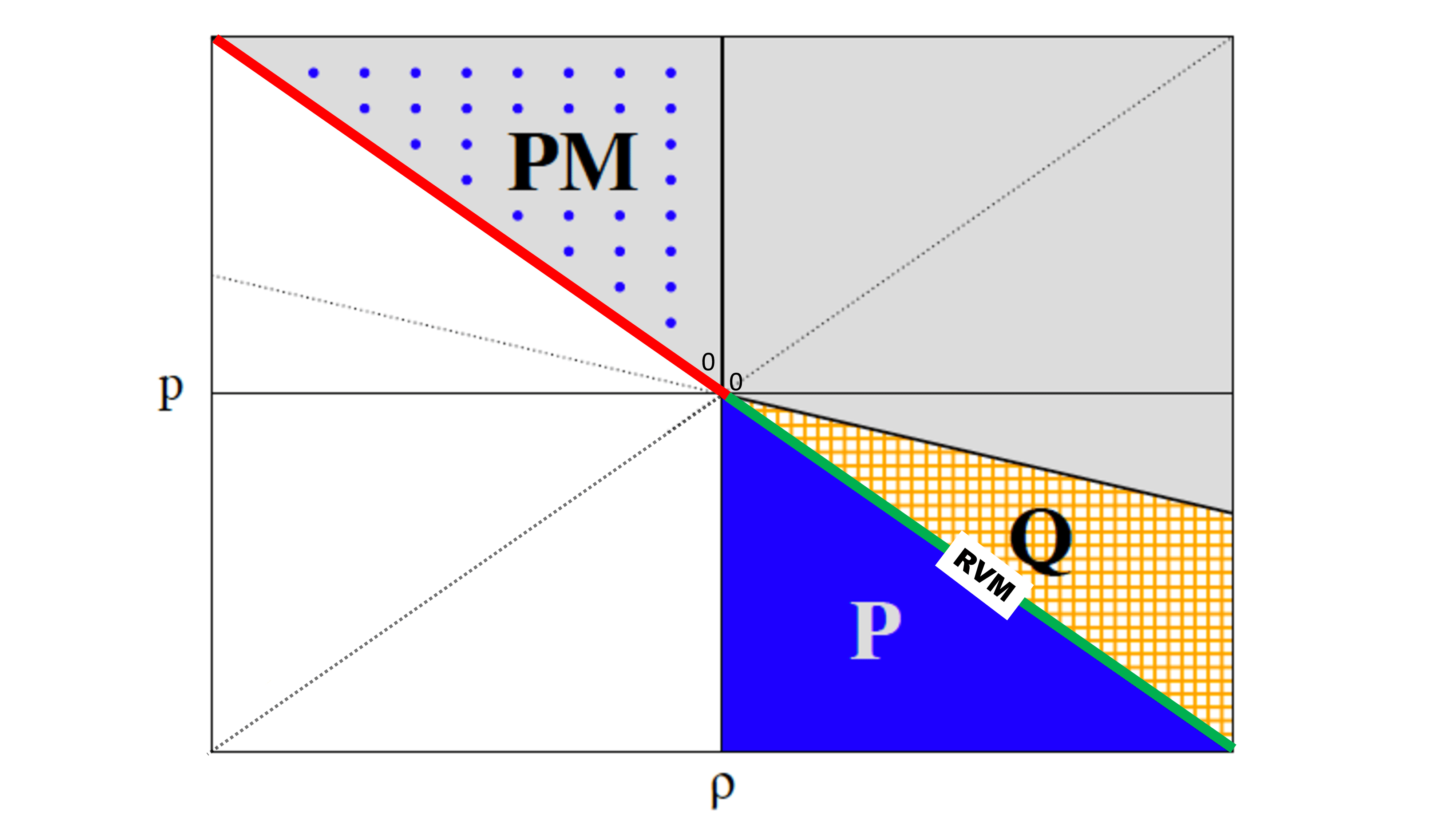}
\end{center}
\caption{\scriptsize EoS diagram for the energy conditions of the cosmic fluids. The quintessence (Q) region ($-1<w<-1/3$ with $\rho>0$, $p<0$) is marked cross-hatched, and the conventional phantom region ($w\le-1$ with $\rho>0$, $p<0$) is the blue sector
indicated as P. The gray-dotted area corresponds to the peculiar  ``Phantom
Matter'' (PM) region:  $w<-1$ with $\rho<0$ and  $p>0$.  Notice that PM satisfies the strong energy condition: $\rho+p\geq 0, \ \rho+3p\geq 0$ (all of  the gray area) {but not the weak energy condition}: $\rho\geq 0, \ \rho+p\geq 0$ (shaded area, {except the P and PM sectors}). The DE component $X$ in our analysis behaves
as PM, whereas the $Y$ component behaves as quintessence. The EoS line $w=-1$ marks off the classical vacuum. The RVM fulfills this EoS only approximately near our time owing to quantum effects\,\citep{Moreno-Pulido:2022upl}. See also \citep{Grande:2006nn}, and \citep{Mavromatos:2021urx} for the modern developments on PM.}
\label{fig:EC}
\end{figure}

Before closing this section and for the sake of a better contextualization of the PM option within the class of energy conditions, in  Fig. \ref{fig:EC} we show a few of the most common possibilities for the EoS of the cosmic fluids.  In it, we particularly highlight the  phantom DE (labelled P,  in blue) and the  phantom matter (PM) regions. The latter (marked gray-dotted)  is far away from the usual phantom DE, it actually lies in its antipodes! These two phantom-like possibilities, both satisfying $w<-1$, are  therefore dramatically different and must not be confused.  It should also be emphasized that while phantom DE has been amply considered in the literature to describe different features of the DE, including a possible explanation for the $H_0$ and growth tensions in different frameworks (see the Introduction and references therein), in our approach phantom DE is not used at all. In point of fact, only PM is singled out in an optimal way within our specific proposal for a possible resolution of the cosmological tensions.


\section{Data and numerical analysis}\label{sec:DataAnalysis}

Let us now come back to the three composite DE models mentioned in Section \ref{sec:CompositeDE}. Leaving model i) for a separate study \citep{LXCDM2024}, in this work we constrain the DE models ii) and iii)  making use of the following cosmological data sets:

\begin{itemize}
\item The full Planck 2018 CMB temperature, polarization and lensing likelihoods \citep{Planck:2018vyg}.

\item The SNIa contained in the Pantheon+ compilation \citep{Brout:2022vxf}, calibrated with the cosmic distance ladder measurements of the SH0ES Team \citep{Riess:2021jrx}. We will refer to this data set as SNIa+SH0ES for short.

\item 33 data points on $H(z)$ in the redshift range $z\in [0.07,1.97]$ from cosmic chronometers (CCH) \citep{Jimenez:2003iv,Simon:2004tf,Stern:2009ep,Zhang:2012mp,Moresco:2012jh,Moresco:2015cya,Moresco:2016mzx,Ratsimbazafy:2017vga,Borghi:2021rft,Tomasetti:2023kek}, see Appendix A of \citep{Favale:2024lgp}. We employ the corresponding full covariance matrix, as described in \citep{Moresco:2020fbm}.

\item Transverse (aka angular or 2D) BAO data from Refs. \citep{Carvalho:2015ica,Alcaniz:2016ryy,Carvalho:2017tuu,deCarvalho:2017xye,deCarvalho:2021azj}. This type of BAO data are claimed to be less subject to model-dependencies, since they are obtained without assuming any fiducial cosmology to convert angles and redshifts into distances to build up the tracer map. The BAO 2D data points are extracted from the two-point angular correlation function or its Fourier transform. In anisotropic (or 3D) BAO analyses, instead, a fiducial cosmology (the standard $\CC$CDM model) is employed to construct the 3D map in redshift space, potentially introducing some model-dependency\footnote{This is an important issue to keep in mind, which is in fact at the basis of our approach in the current presentation.  It is known that the impact of the fiducial cosmology on the BAO measurements can be small \citep{Carter:2019ulk,Heinesen:2019phg,Bernal:2020vbb,Pan:2023zgb}, but this is warranted only when such a cosmology is close to the true one. In fact, \cite{Anselmi:2022exn} recently argued that these works do not explore a wide enough range of the fiducial parameter values allowed by the measured BAO distances, which are quite far away  from the Planck best-fit $\Lambda$CDM model. In addition, it is claimed that by fitting the two-point correlation function while fixing at the same time the cosmological parameters (and also those entering the correlation
function template) at fiducial values, might lead to an underestimation of the errors by roughly a factor of two, hence a rather significant potential effect \citep{Anselmi:2018vjz}.}. It is also worth noticing that despite the fact of being constructed  from the same parent catalogues of tracers,
 angular and anisotropic BAO data exhibit some degree of tension, see e.g. \citep{Camarena:2019rmj} and, particularly, the very recent analysis of \cite{Favale:2024sdq}. In the last study, it is shown that  upon excluding the radial component of BAO 3D (which is of course not present in BAO 2D) the `residual' tensions left between the two BAO types can still attain in between $2\sigma$ to $4.6\sigma$, depending on the data set used. All in all, not surprisingly, these BAO data tensions may directly impinge in a significant way on the discussion of the cosmological tensions themselves, for the required solution to alleviate them may be conditioned to the specific BAO data type considered in the analysis \citep{Gomez-Valent:2023uof}.  We should also mention that, in contradistinction to the BAO 3D data, BAO 2D observations still offer room  for low-redshift solutions to the Hubble tension while respecting the constancy of the absolute magnitude of SNIa \citep{Akarsu:2023mfb,Gomez-Valent:2023uof}. Given, however, the persisting conflict between these two BAO types, our aim here is to remain as model-independent as possible in the light of the present knowledge, and for this reason we opt for using at this point the kind of BAO data which seem to be less subject to criticism at present (and in this sense that may be more trouble-free),  to wit: the BAO 2D type. We leave the detailed comparison with the BAO 3D results for a more comprehensive exposition. A preview of such a comparative analysis will be nonetheless advanced in the conclusions (see Section \ref{sec:Conclusions})\footnote{The extended discussion, including an explicit comparison between the results obtained from  BAO 2D and BAO 3D data, as  well as the computational details of the present work, will be disclosed in a forthcoming comprehensive study \citep{LXCDM2024}.}.

\item The data on large-scale structure (LSS) at $z\lesssim 1.5$ from Refs. \citep{Guzzo:2008ac,Song:2008qt,Blake:2011rj,Blake:2013nif,Simpson:2015yfa,Gil-Marin:2016wya,eBOSS:2020gbb,Said:2020epb,Avila:2021dqv,Mohammad:2018mdy,Okumura:2015lvp}. These are data points on the observable
 $f(z)\sigma_8(z)$, with $f(z)=-(1+z)d\ln\delta_m/dz$ the growth rate, $\delta_m=\delta\rho_m/\rho_m$ the matter density contrast, and $\sigma_8(z)$ the rms mass fluctuations at a scale $R_8=8h^{-1}$ Mpc, see Table 3 of  \citep{SolaPeracaula:2023swx}. These data points, though, are taken by the observational groups using a fiducial cosmology with $h\sim 0.67$, which translates into measurements at a characteristic scale of $R\sim 12$ Mpc. Here, in contrast, we adhere to the reasoning and practice of Refs. \citep{Sanchez:2020vvb,eBOSS:2021poy,Semenaite:2022unt}, and we treat these observations as data points on $f(z)\sigma_{12}(z)$, using a Fourier-transformed top-hat window function $W(kR_{12})$  in the computation of $\sigma_{12}(z)$. {The advantage, as emphasized by these authors,  is that the scale $R_{12}=12$ Mpc is independent of the parameter $h$. }
\end{itemize}

We compute all the cosmological observables using a modified version of the Einstein-Boltzmann code CLASS \citep{Lesgourgues:2011re,Blas:2011rf}, and   explore and constrain the parameter space of the various models using the Metropolis-Hastings algorithm  \citep{Metropolis:1953am,Hastings:1970aa} implemented in MontePython \citep{Audren:2012wb,Brinckmann:2018cvx}. The resulting Monte Carlo Markov chains are analyzed with the Python code GetDist \citep{Lewis:2019xzd}. Our main numerical results are presented in Table \ref{tab:table_fits} and in the triangle plots of Figs. \ref{fig:triangle_plot_in}-\ref{fig:wXwY}. Supplementary information on plots and tables is provided in the Appendix, which will be commented on in the main text.

\begin{table*}[t!]
\centering
\renewcommand{\arraystretch}{2}
\resizebox{\textwidth}{!}{
\begin{tabular}{|c ||c | c  |c |}
\hline
{\small Parameter} & {\small $\Lambda$CDM}   & {\small $w$XCDM }  & {\small $\Lambda_s$CDM}
\\\hline
$\omega_b$ & $0.02281\pm 0.00014$ (0.02278) & $0.02241\pm 0.00013$ (0.02260) & $0.02236^{+0.00016}_{-0.00018}$ (0.02232) \\\hline
$\omega_{\rm dm}$ &  $0.1153\pm 0.0009$ (0.1148)   & $0.1199\pm 0.0010$ (0.1196) & $0.1205^{+0.0015}_{-0.0016}$ (0.1216) \\\hline
$\ln(10^{10}A_s)$ & $3.066^{+0.016}_{-0.018}$ (3.080) &  $3.037\pm 0.014$ (3.034)  & $3.036^{+0.015}_{-0.016}$ (3.030) \\\hline
$\tau$ & $0.069^{+0.008}_{-0.010}$ (0.076)  & $0.051\pm 0.008$ (0.048) &  $0.050^{+0.008}_{-0.009}$ (0.046)   \\\hline
$n_{s}$ &  $0.978\pm 0.004$ (0.981) &   $0.967\pm 0.004$ (0.969) &  $0.966^{+0.004}_{-0.005}$ (0.961)  \\\hline
$H_{0}$ [km/s/Mpc] &  $69.82^{+0.41}_{-0.44}$ (70.05) &  $72.75^{+0.57}_{-0.71}$ (72.36) & $72.24^{+0.99}_{-0.75}$ (73.82)\\\hline
$z_t$ & $-$ & $1.46^{+0.02}_{-0.01}$ (1.47) & $1.61^{+0.22}_{-0.18}$ (1.47) \\\hline
$w_X$ & $-$  & $-1.16^{+0.13}_{-0.16}$ (-1.16)  &  $-$ \\\hline
$w_Y$ & $-$  & $-0.90\pm 0.03$ (-0.88)&  $-$ \\\hline\hline
$\Omega_m^0$ & $0.283\pm 0.005$ (0.280) & $0.269\pm 0.005$ (0.272)  &  $0.267\pm 0.005$ (0.264)\\\hline
$M$ &  $-19.372^{+0.011}_{-0.012}$ (-19.362)   & $-19.273^{+0.015}_{-0.016}$ (-19.282) & $-19.278^{+0.026}_{-0.020}$ (-19.261)   \\\hline
$\sigma_{12}$ & $0.780\pm 0.007$ (0.784) &  $0.776\pm 0.007$ (0.772) & $0.782^{+0.007}_{-0.006}$ (0.784)
 \\\hline\hline
$\chi^2_{\rm min}$ &  $4166.76$  & $4107.62$ &  $4120.04$ \\\hline
$\Delta$DIC &  $-$  & $57.94$  & $40.16$  \\\hline
$\Delta$AIC &  $-$  & $53.14$  & $44.72$\\\hline
\end{tabular}}
\caption{\scriptsize Mean values and uncertainties at 68\% CL obtained with the full data set CMB+CCH+SNIa+SH0ES+BAO+$f\sigma_{12}$. We show the best-fit values in brackets. We use the standard notations for the $\CC$CDM parameters. In the last three lines, we display the values of the minimum $\chi^2$, $\Delta$DIC and $\Delta$AIC, as defined in Eqs. \eqref{eq:DIC} and \eqref{eq:AIC}, respectively.  Positive values of $\Delta$DIC and $\Delta$AIC denote preference of the new models over the $\CC$CDM.  As can be seen, the preference is extraordinarily  high.}
\label{tab:table_fits}
\end{table*}

In Table \ref{tab:table_fits} we do not only list the mean of the various parameters, with their uncertainties and best-fit values, but also report the minimum values of $\chi^2$ obtained for each model and the difference between the deviance (DIC) \citep{DIC} and Akaike (AIC) \citep{Akaike} information criteria found between the composite DE models and $\Lambda$CDM, which we treat as our fundamental benchmark model. These differences read $\Delta{\rm DIC}\equiv{\rm DIC}_{\Lambda{\rm CDM}}-{\rm DIC}_{i}$ and $\Delta{\rm AIC}\equiv{\rm AIC}_{\Lambda{\rm CDM}}-{\rm AIC}_{i}$, respectively, with $i$ referring to $w$XCDM or $\Lambda_s$CDM. Both criteria penalize the use of additional parameters and can be regarded as a rigorous mathematical implementation of Occam's razor. The DIC is defined as

\begin{equation}\label{eq:DIC}
   {\rm DIC} =  \chi^2(\bar{\theta})+2p_D\,,
\end{equation}
with $p_D=\overline{\chi^2}-\chi^2(\bar{\theta})$ the effective number of parameters in the model, $\overline{\chi^2}$ the mean value of $\chi^2$, and $\bar{\theta}$ the mean of the parameters that are left free in the Monte Carlo analysis. Similarly, AIC is defined as

\begin{equation}\label{eq:AIC}
    {\rm AIC} =  \chi^2_{\rm min}+2n_p\,,
\end{equation}
with $n_p$ the number of free parameters entering the fit. We note that the above formula is a good approximation when the number of data points entering the fit is much larger than $n_p$, which is certainly the case here.  With our definition, a positive difference of these information criteria implies that the composite DE models perform better than the $\Lambda$CDM, whereas negative differences imply just the opposite. According to the usual jargon of the information criteria, if $0 \leq \Delta\textrm{DIC}<2$ it is said that one finds \textit{weak evidence} in favor of the new model under test (in this case, the composite DE models under scrutiny), as compared to the standard model. If  $2 \leq \Delta\textrm{DIC} < 6$ we speak, instead, of \textit{positive evidence}. If $6 \leq \Delta\textrm{DIC} < 10$, there is \textit{strong evidence} in favor of the composite DE models, whilst for $\Delta\textrm{DIC}>10$ we may legitimately conclude that there is \textit{very strong} statistical evidence supporting the new model or models against the standard $\CC$CDM. Analogous considerations can be made using AIC, of course. DIC is considered to be a more accurate information criterion, since it incorporates the information encapsulated in the full Markov chains. However, for the sake of generality and to explicitly show the consistency between these two criteria, we display the results obtained from both criteria at the same time in our Table \ref{tab:table_fits}.

\section{Discussion}\label{sec:Discussion}

As we have mentioned, bubbles of PM are connected with the process of attaining the de Sitter (dS) era both in the early and in the late universe. During the transit from the provisional phantom vacuum  into the true and stable (dS) vacuum, the PM bubbles being tunneled in the late universe are characterized by positive pressure and hence offer a fertile field for the growing of unsuspected structures.  This mechanism, therefore,  provides a tantalizing theoretical framework capable of explaining physically the appearance of a late time AdS epoch (or epochs) before the usual dS epoch is eventually attained \footnote{This is in contradistinction to the alternative $\CC_s$CDM scenario analyzed in \citep{Akarsu:2023mfb},  although theoretical attempts at justifying it have appeared recently \citep{Anchordoqui:2023woo,Akarsu:2024qsi,Akarsu:2024eoo,Anchordoqui:2024gfa}.}.  In point of fact, we do not expect that these PM bubbles are perfect AdS phases (nor that the subsequent stage is necessarily characterized by a  rigid $\CC>0$) and for this reason we have left the EoS parameter $w_X$ as a free parameter, whose fitted value turns out to be $w_X\simeq -1.16$ (cf. Table \ref{tab:table_fits}), although with $\rho_X<0$ (in contrast to conventional phantom DE). Since the bubble is transitory, the ensuing DE phase below  $z_{t}$ carries another EoS which we fit it to be quintessence-like: $w_Y\simeq -0.90$ (see  Table \ref{tab:table_fits}).  We have performed a comparative fitting analysis of the $w$XCDM and $\CC_s$CDM scenarios using the same data sets and have found a significantly better fit quality for the PM scenario, as shown in Table \ref{tab:table_fits}. We shall come back to these results in a moment.

The PM phase (or phases) with negative energy density and positive pressure are left behind during the cosmic evolution at relatively large redshifts of order $1-10$, and in these places they can leave a sort of oasis rich of large scale structures, whereas the consecutive evolution (closer to our time) flips into the quintessence regime.  Although we have used an abrupt $\theta$-function behavior to connect the two EoS regimes, in actual fact the process is continuous since the Chern-Simons condensates eventually dominate and restore the normal vacuum phase with positive energy. To better understand, in a quantitative way, how the PM bubbles with positive pressure may enhance the formation of (unsuspected) large scale structures in the relatively distant past, it is useful to analyze the differential equation for the matter density contrast in the presence of PM. This is a key aspect of the role played by PM for potentially helping to solve the cosmological tensions. Before the transition at $z_{t}$ (i.e. for $z>z_{t}$), the equation for $\delta_m$ reads\footnote{As indicated, we shall not provide computational details here. For a full fledged exposition, including the complete set of coupled perturbations equations of the \wXCDM model, see \citep{LXCDM2024}.}
\begin{equation}\label{eq:dcX}
\delta^{\prime\prime}_m+\frac{3}{2a}\left(1-\Omega_X(a)w_X\right)\delta^\prime_m-\frac{3}{2a^2}(1-\Omega_X(a))\delta_m=0\,,
\end{equation}
with the primes denoting derivatives with respect to the scale factor. PM has negative energy density ($\Omega_X<0$) and positive pressure (due to $w_X<-1<0$) and, therefore, it induces a decrease of the friction term and an increase of the Poisson term (the last one) in Eq. \eqref{eq:dcX}. Both effects are in harmony and therefore bring about, given some fixed initial conditions, a net enhancement of the structure formation processes in the PM bubbles. Notice that this would \textit{not} occur for ordinary phantom DE nor for quintessence (cf. Fig.\,\ref{fig:EC}), for which the friction term gets enhanced and the Poisson term suppressed, i.e. just the opposite situation to PM. This conspicuous  impact on LSS is therefore unique to PM since there is no other cosmic fluid capable of matching such an achievement in the EoS diagram of Fig.\,\ref{fig:EC}.

\begin{figure}[t!]
    \centering
    \includegraphics[scale=0.13]{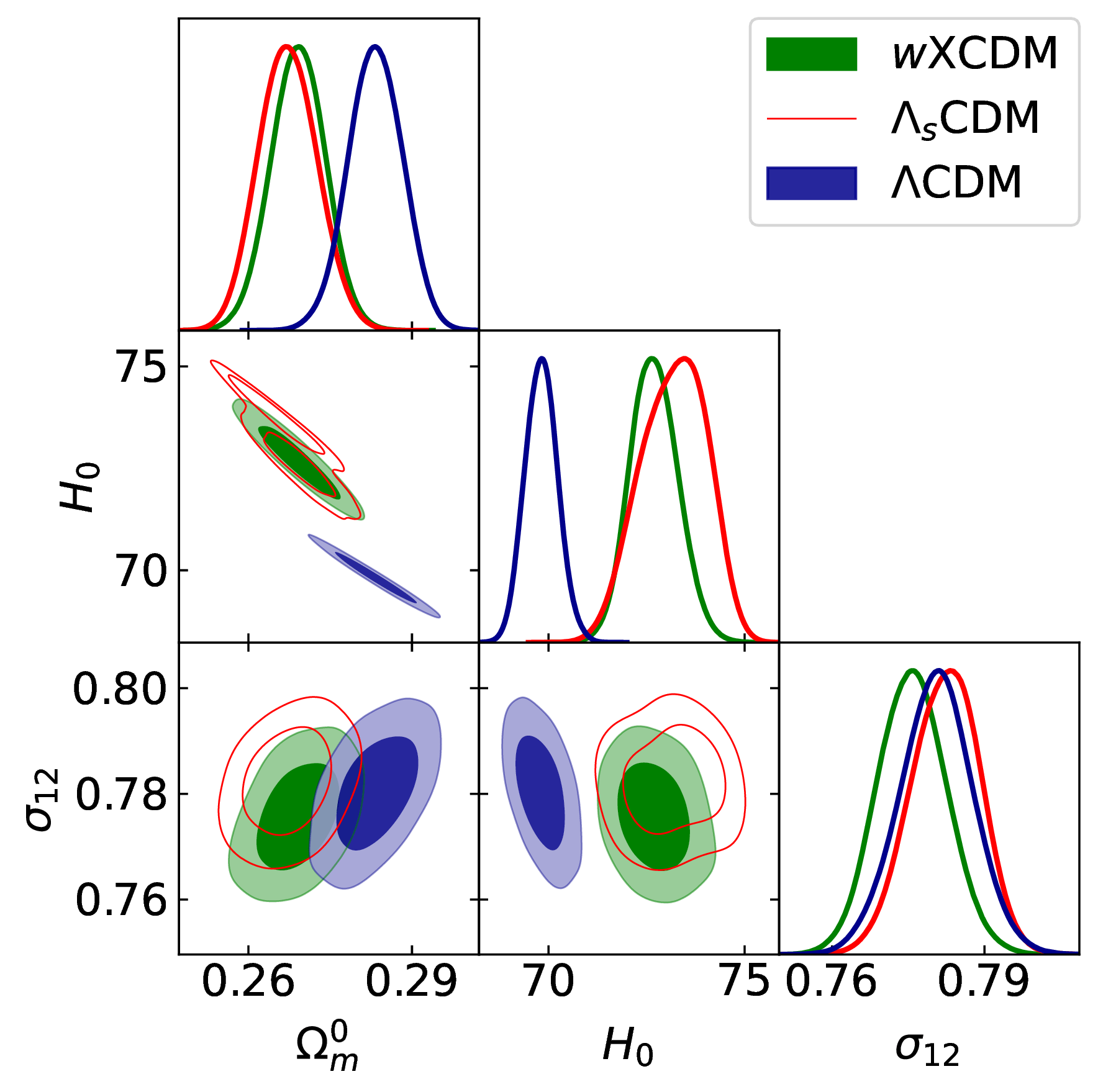} \caption{\scriptsize Contour plots at $68\%$ and $95\%$ CL and the corresponding one-dimensional posterior distributions for some of the parameters that are relevant in the discussion of the cosmological tensions, for all the models under study. $H_0$ is given in km/s/Mpc. The complete triangle plot is presented in Fig. \ref{fig:full_triangle_plot} of the Appendix.}
    \label{fig:triangle_plot_in}
\end{figure}

Moreover, and this is crucial to understanding how the $H_0$ tension can be potentially cured in the PM framework: because the energy of the $X$ entity is negative, $\Omega_X<0$ (and non-negligible at high redshift), it enforces a higher value of the expansion rate $H$ in the quintessence stage in order to preserve the angular diameter distance to the {last scattering surface, which is essentially fixed from the very precise measurement of  $\theta_*$ (the angular size of the sound horizon)  by Planck \citep{Planck:2018vyg} and the standard physics before recombination \citep{Gomez-Valent:2023uof}}. This explains on plausible physical grounds why $H_0$ is found larger than in the $\CC$CDM in our PM scenario. Quantitatively, our value of $H_0$ (cf. Table \ref{tab:table_fits}) is in full agreement with that of  SH0ES ($H_0=73.04\pm 1.04$ km/s/Mpc) \citep{Riess:2021jrx} to within $\lesssim 0.25\sigma$. The Hubble tension is therefore virtually washed out, also when it is formulated in terms of the absolute magnitude of SNIa:  our fitted value of $M$ and that of SH0ES ($M=19.253 \pm 0.027$ mag) differ by only $\sim 0.6\sigma$. At the same time the rate of LSS formation becomes suppressed below $z_{t}\sim 1.5$ during the quintessence-like regime, as explained above, which is also in accordance with the observations. In Table \ref{tab:table_fits} and, more graphically, in Fig. \ref{fig:triangle_plot_in}, we can see that the amplitude of the power spectrum at linear scales that is preferred by the data is pretty similar in all the models under study. Indeed, in all cases we find values of $\sigma_{12}\sim 0.78$. However, at a finer level of scrutiny,  Table \ref{tab:table_chi2} in the Appendix tells us that the $w$XCDM is able to describe better the LSS data than the $\Lambda$CDM, what means that the composite DE model is able to produce lower values of $f(z)\sigma_{12}(z)$ and, hence, of $f(z)$ since $\sigma_{12}$ remains stable in the various models. This can again be understood by looking at the equation for the density contrast, whose form at $z<z_{t}$ is identical to Eq. \eqref{eq:dcX}, but with the replacements $\Omega_X\to \Omega_Y$ and $w_X\to w_Y$,

\begin{equation}
\delta^{\prime\prime}_m+\frac{3}{2a}\left(1-\Omega_Y(a)w_Y\right)\delta^\prime_m-\frac{3}{2a^2}(1-\Omega_Y(a))\delta_m=0\,.
\end{equation}
It is clear from Fig. \ref{fig:triangle_plot_in} that the values of the matter density parameter $\Omega_m^0=\Omega_m(a=1)$ in the $w$XCDM are a way smaller than in $\Lambda$CDM. This obviously translates into larger values of $\Omega_Y(a)>0$ (higher than $\Omega_\CC^0$ in the $\CC$CDM), which makes matter fluctuations to grow less efficiently in our model after the transition (at $z<z_{t}$), a welcome feature that is rightly aligned for potentially solving also the growth tension within the PM picture. As previously mentioned, despite the presence of PM in the past, the value of $\sigma_{12}$  in  $w$XCDM is not significantly different ($\sim 0.4\sigma$) from that of $\Lambda$CDM. This is due to the slower increase of matter fluctuations in the late-time universe and the smaller value of the amplitude of primordial fluctuations ($A_s$)  found in $w$XCDM, which somehow compensate for the presence of PM at $z>z_t$ an keep the value of $\sigma_{12}$ stable (cf. again Table \ref{tab:table_fits}, and Fig. \ref{fig:full_triangle_plot} in the Appendix).

From our numerical analysis we can see that both models $w$XCDM and  $\Lambda_s$CDM  offer a dramatic reduction of the cosmological tensions. This is apparent from Table \ref{tab:table_fits}. One could say that the Hubble and growth tensions disappear in these models. However, there are also important quantitative (and conceptual)  differences between them.  In terms of the information criteria, we find that the two models  $w$XCDM and  $\Lambda_s$CDM are very strongly preferred over $\Lambda$CDM.  Numerically, however,  the  $w$XCDM solution provides a significantly better global fit.  For this model we find $\Delta$DIC,$\Delta$AIC$\gtrsim 55$, whereas for $\Lambda_s$CDM we obtain $\Delta$DIC,$\Delta$AIC$\gtrsim 40$. The difference of $15$ units is quite substantive and indicates (in the common parlance of the information criteria) that there is very strong evidence that model $w$XCDM is preferred over model  $\Lambda_s$CDM (for the same data).   Worth noticing is also the fact that our own results for $\Lambda_s$CDM agree with those originally reported in \citep{Akarsu:2023mfb}, despite the existing differences in the data sets employed in the two analyses. Apart from our using data on CCH, which were not considered in \citep{Akarsu:2023mfb}, we employ data on $f(z)\sigma_{12}(z)$ instead of the weak lensing (WL) data from KiDS. We do not use WL data in our analysis in order not to compromise the computation of the non-linear matter power spectrum, which might not be sufficiently well under control in the models under study. Our fitting results, though, are completely consistent with theirs, leading also to similar values of $S_8\sim 0.78$ (as we have checked), the latter being a more natural LSS parameter in the context of WL analyses, but not in our approach.

Perhaps the most remarkable outcome of our analysis is that, in the light of the overall fit quality of our results, which lead to such an outstanding support to the composite DE models from the information criteria,  it should be fair to conclude that the standard $\Lambda$CDM model appears to be comparatively ruled out at a very high confidence level. However, once more, we have to warn the reader that, in principle,  such a conclusion ensues under the assumption that only BAO 2D have been employed in our analysis (see, however, the additional comments in the Conclusions, i.e. Section \ref{sec:Conclusions}).
In an attempt to further understand the great success of the numerical fits from the composite DE models in Table \ref{tab:table_fits}, let us analyze the degree of impact from each data source. The quantitative influence of each data source  can be inferred directly from Table \ref{tab:table_chi2} in the Appendix, where a detailed breakdown of the different contributions to $\chi^2$ is reckoned. The angular/transversal BAO data certainly contributes in a significant way, as could be expected from the analysis of \citep{Gomez-Valent:2023uof}. However, it is by no means the leading contribution, as it is  responsible for roughly $\sim 18\%$ of the success.  A close inspection of the mentioned table shows that the composite DE models beat actually the $\Lambda$CDM in every single observable, and not only in those driving the tensions. Particularly noticeable is the sizeable impact from such a solid asset of basic data as the CMB and SNIa+SH0ES observations, which are responsible for about $70\%$ of our successful fitting result, whereas the influence from LSS data is less than $10\%$; and, finally, that of  CCH is marginal (a few percent). The bulk of the successful fit, therefore, relies on the most fundamental cosmological data, which is  remarkable.  The $\Lambda$CDM is not only unable to explain the large value of $H_0$ measured by SH0ES or to further slow down the growth rate at low redshift, but it also introduces other tensions, e.g. with the value of the (reduced) cosmological matter parameter  $\omega_{\rm m}=\omega_{b}+\omega_{\rm dm}$ that is preferred by the Planck data in models which assume standard model physics before the decoupling of the CMB photons, $\omega_m=0.142\pm 0.001$ \citep{Planck:2018vyg}. The latter is fully consistent with the $w$XCDM value ($\omega_m= 0.142\pm 0.001$), but differs by $2.8\sigma$ with the one inferred in the standard $\Lambda$CDM ($\omega_m=0.138\pm 0.001$). All that said, we remark that the BAO data play indeed a momentous role in the $H_0$ tension, for  despite the fact that the \textit{overall} fit with  composite DE does substantially improve using any of the two variants of BAO, the specific relieve of the $H_0$ tension hinges dramatically upon the particular sort  (2D or 3D)  of BAO that is employed, see Section \ref{sec:Conclusions}.

We should emphasize that the global outperformance of the composite DE models under study as compared to the $\CC$CDM is obtained after duly penalizing the use of extra parameters in them.  This is of course implemented automatically by the information criteria, as explained above.  Thus, for the $\CC_s$CDM there is the transition redshift  $z_{t}$ as a new parameter with respect to the $\CC$CDM, whereas for the $w$XCDM we have, in addition, the two EoS parameters  $w_X$ and $w_Y$ for the two phases of the DE. Despite the presence of these extra parameters, Occam's razor (formalized through the verdict of the information criteria) still bestows exceptional preference for the composite models over the concordance model. This is the first important conclusion of our analysis, which suggests that the composite nature of the DE could be a fact. The second, is that the relative differences between the two composite models under scrutiny, namely DIC$_{\Lambda_s\rm CDM}$-DIC$_{w\rm XCDM}=17.78$ and AIC$_{\Lambda_s\rm CDM}$-AIC$_{w\rm XCDM}=8.42$, clearly point to a rather keen preference for $w$XCDM over $\Lambda_s$CDM under the same data.

\begin{figure}[t!]
    \centering
    \includegraphics[scale=0.3]{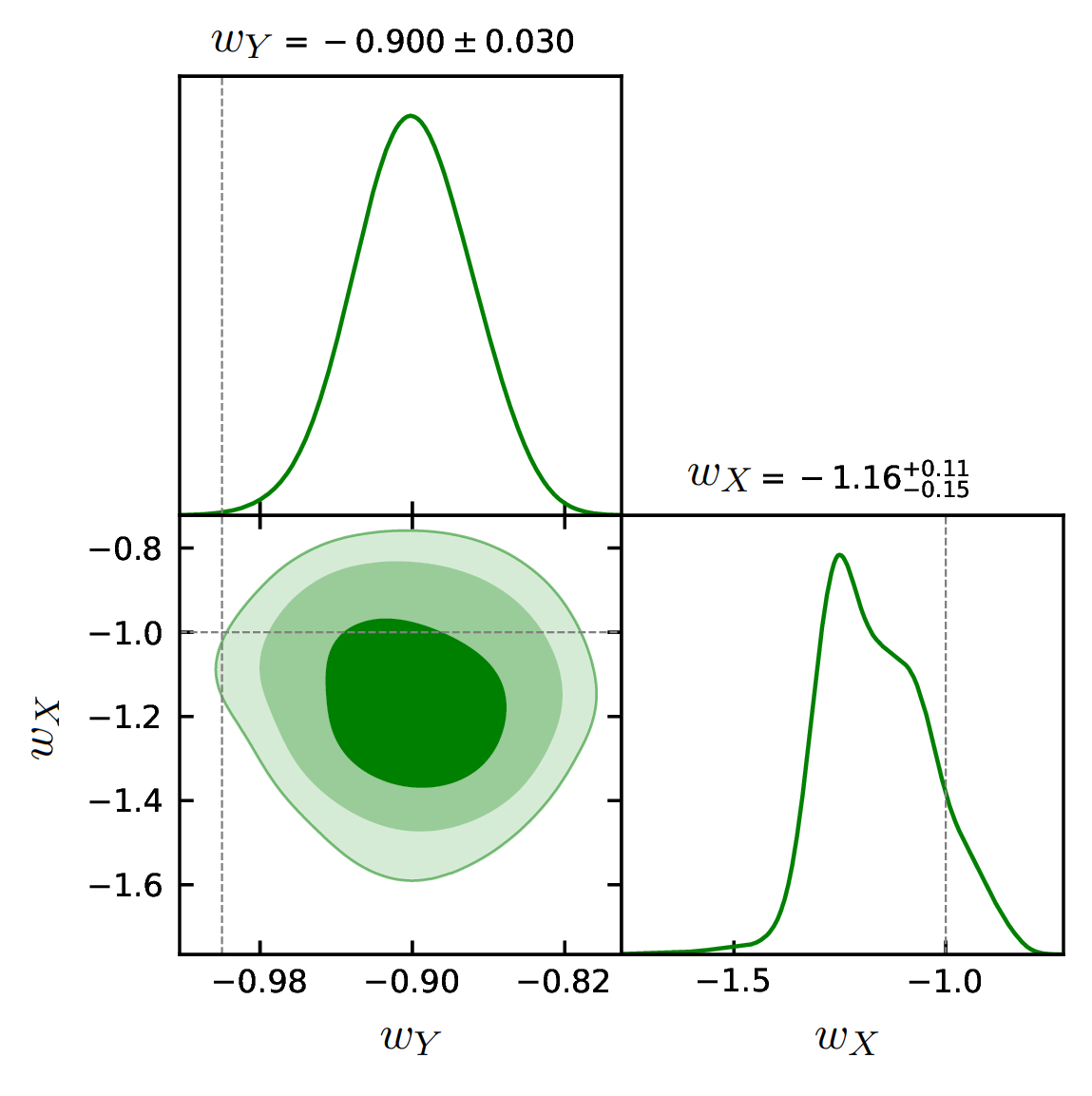}
    \caption{\scriptsize Confidence regions in the $w_Y-w_X$ plane of the $w$XCDM model, and the corresponding one-dimensional posterior distributions. The dotted lines are set at $w_Y=-1$ and $w_X=-1$. The intersection of the horizontal and vertical lines in the contour plot corresponds to the $\Lambda_s$CDM model, which falls $\gtrsim 3\sigma$ away from the preferred region of the $w$XCDM. See the main text for further comments.}
    \label{fig:wXwY}
\end{figure}

Focusing now on  the $w$XCDM parameters, in Fig. \ref{fig:wXwY} we show the constraints obtained in the EoS plane  $w_Y$-$w_X$, which involves two of the characteristic new parameters of the $w$XCDM model (the third one being $z_t$, shared with $\CC_s$CDM). The central value of $w_X=-1.16$ falls in the phantom region (in fact, PM region since $\Omega_X<0$) but is compatible with $-1$ at $\sim 1\sigma$ c.l.  There is, in contrast, a non-negligible ($\sim 3.3\sigma$) preference for a quintessence-like evolution of the  DE for the low redshift range nearer to our time ($z<z_{t}$): $w_Y=-0.90\pm 0.03$ (see also Fig.\,\ref{fig:wXwY}). {The preference that we find for quintessence-like behavior in this region can be contrasted with the situation in the $\CC_s$CDM, where a rigid $\Lambda$-term is assumed below the transition redshift (also above that redshift, with $\Lambda\to -\Lambda$). Now the fact that our fit with model $w$XCDM  proves significantly better than with model $\CC_s$CDM  (as commented above) suggests that indeed the quintessence option in the low redshift range is strongly preferred. Our result aligns perfectly well with} the one obtained from the analysis of the Pantheon+ data in the context of the flat $w$CDM parametrization \citep{Brout:2022vxf}, which showed that SNIa data alone lead to an EoS parameter $w=-0.89\pm 0.13$ after marginalizing over $\Omega_m^0$. {This marginalized result, though,  is still compatible with a CC ($w=-1$). To explain why in our case is more focused on the quintessence domain, let us note that } for $w$XCDM we find a tight constraint on the matter density parameter, strongly favoring the region of small values, $\Omega_m^0=0.269\pm 0.005$. This is induced by two facts: (i) CMB prefers values of $\omega_m\sim 0.142$ in models with standard pre-recombination physics; and (ii) the large values of $H_0\sim 73$ km/s/Mpc measured by SH0ES. Combining the aforementioned tight constraint on $\Omega_m^0$ with the constraints in the $w-\Omega_m^0$ plane obtained for the $w$CDM from the analysis of the Pantheon+ data alone (cf. the contours in cyan of Fig. 9 of \citep{Brout:2022vxf}) we can break the existing degeneracy in that plane and explain the $\gtrsim 3\sigma$ evidence for quintessence DE that we find within the $w$XCDM for the low redshift range. We point out that, for the conventional $w$CDM parameterization (i.e. without the $X$ component),  such low values of $\Omega_m^0$ are not favored by Planck, since they are in tension with the angular diameter distance to the last scattering surface, see again Fig. 9 of \citep{Brout:2022vxf}. Remarkably, this problem can be fully averted in the $w$XCDM model owing precisely  to the concurrence of the $X$ component, which can act as PM in the high redshift stretch $z>z_{t}$ and compensate for the decrease of the contribution to the angular diameter distance in the low redshift domain. On the other hand, in the $\Lambda_s$CDM case we also find similar low values of $\Omega_m^0$. However,  to retrieve this model from the $w$XCDM we have to set $w_Y$ to $-1$ (rigid CC)\footnote{Note that the $\Lambda_s$CDM model can be viewed as just the single point $(w_Y,w_X)=(-1,-1)$ in the \wXCDM parameter space of  Fig.\,\ref{fig:wXwY}.}, so the $\CC_s$CDM falls outside the $3\sigma$ region obtained with $w$CDM using only SNIa data \citep{Brout:2022vxf}. This issue is also solved in the $w$XCDM and explains the improvement found with respect to the $\Lambda_s$CDM model. The full 5-year data set of high-redshift SNIa from the Dark Energy Survey  \citep{DES:2024tys}  and the Union3 SNIa compilation \citep{Rubin:2023ovl} will most likely increase the aforesaid signal, since they also find constraints in the region of interest, which read $(\Omega_m^0,w)=(0.264^{+0.074}_{-0.096},-0.80^{+0.14}_{-0.16})$ and $(\Omega_m^0,w)=(0.244^{+0.092}_{-0.128},-0.735^{+0.169}_{-0.191})$, respectively. Hence, it will be of utmost importance to study the impact of these data sets in future analyses. In addition, our fitting results for $w$XCDM are compatible at  roughly $\sim 1\sigma$ with those recently reported by DESI  using the $w$CDM parametrization\,\citep{DESI:2024mwx}. Although two of the twelve DESI BAO data points (specifically those obtained with Ly-$\alpha$ quasars (QSO) at $z_{\rm eff}=2.33$)  lie outside the quintessence region identified in our $w$XCDM model, the bulk of the DESI points lie below $z=1.5$. The exclusion of the Ly-$\alpha$ QSO data points in their fitting analysis might even improve further the compatibility with our findings. Additionally,  some recent analyses using DESI data to reconstruct the dark energy density have found hints of negative values at $z>1.5$ \citep{DESI:2024aqx,Orchard:2024bve}. Since the DESI data are derived from anisotropic (3D) BAO, exploring future angular BAO data from DESI (if available) and comparing them with the BAO 3D data would be an interesting avenue for further research.

\section{Conclusions}\label{sec:Conclusions}

In this paper, we have addressed a possible solution to the cosmological tensions within a simplified version of the old existing $\CC$XCDM framework\,\citep{Grande:2006nn,Grande:2006qi,Grande:2008re},  which was born in the theoretical arena of the RVM -- see \citep{SolaPeracaula:2022hpd,Sola:2013gha} and references therein. Our modern approach benefits  from the new theoretical developments in the context of the stringy version of the running vacuum model (StRVM) \citep{Mavromatos:2021urx,Mavromatos:2020kzj}.  The entity $X$ that is involved in the $\CC$XCDM need not be a fundamental field but can display a phantom matter (PM) behaviour, which in the StRVM is predicted to appear in the cosmic transit to the de Sitter phase. The PM regime is, therefore,  only transitory and acts as a temporary stage with negative energy and positive pressure. It can be a pure anti-de Sitter (AdS) phase but in general its EoS $w_X$ is not necessarily $-1$, although it must satisfy $w_X\lesssim -1$. In this work we have contented ourselves with a simplified version of the full $\CC$XCDM, which we have called the $w$XCDM. We have utilized a robust data set, which includes the Pantheon+ compilation of SNIa, cosmic chronometers, transverse BAO, large-scale structure data, and the full CMB likelihood from Planck 2018. As warned previously, our exclusive use of transverse (2D) BAO  aims at maximally mitigating the model-dependent effects that may be introduced on using anisotropic (3D) BAO, according to some published studies. No similar issues have been identified in BAO 2D and hence we have focused here on using only uncontested BAO data, with the hope that the situation with BAO 3D will be completely clarified in the near future (see, however, the additional comments below).

Using the $w$XCDM and the aforementioned data configuration}, we find that the PM phase shows up above a transition redshift $z_{t}\simeq 1.5$ (cf. Table \ref{tab:table_fits}). For $z>z_{t}$ the PM behavior is rapidly washed out since its energy density {fraction} becomes smaller and smaller in the past. And, on the other hand, its effect towards our time ($z<z_{t}$) becomes erased by the progressive appearance of the Chern-Simons condensates, which are responsible for eventually establishing a steady $\CC=$const.$>0$  regime \citep{Mavromatos:2021urx}. For this reason, in practice, the appearance of a PM phase is confined to a bubble of  spacetime, which is left behind. However, in that space and time new structures can emerge, being completely  unsuspected in the context of the standard model. The PM bubbles that are met within the stringy RVM formulation originate as a result of a  ``phantom vacuum'' phase, which emerges in the universe when the expansion heads towards the next de Sitter epoch.  While the implications of this fact for the early universe were considered previously \citep{Mavromatos:2021urx}, here we have extended this phenomenon to the late universe.  In both cases the PM bubbles are localized spacetime events. They can only occur as tunneling attempts to establish the phantom vacuum \citep{Mavromatos:2021urx} with positive pressure and negative energy: $p=-\rho>0$.  Below the transition redshift, the PM phase ceases to occur and the field $Y$ (viz. the would-be running VED in the full $\CC$XCDM\,\citep{Grande:2006nn})  takes its turn  with a new effective EoS, which is fitted to be of quintessence type ($w_Y\gtrsim -1$) at $\sim 3.3\sigma$ c.l. As a matter of fact, this is the effective form of DE that is expected in our most recent past ($z<1.5$) within the $w$XCDM, although from a more fundamental level it could just be running vacuum energy with $\nueff>0$ in the full $\CC$XCDM model, see Eq. \eqref{eq:RVM}. The observed DE should thereupon be dynamical, and this important conclusion is in stark contrast with the alternative $\CC_s$CDM scheme analyzed in\,\citep{Akarsu:2023mfb}, in which below the transition redshift there is a {rigid and positive cosmological term $\CC=$const.  Actually, no possible time dynamics for the DE is available for the $\CC_s$CDM, neither above nor below $z_t$. On this account,  the picture emerging from the $w$XCDM looks more consistent with the first data release by DESI \citep{DESI:2024mwx}, which points to dynamical DE. As noted in the Introduction, the fashionable conclusion about potential evidence of time-evolving DE is actually well in consonance with previous phenomenological studies of the RVM from several years ago, which were based on a pretty large set of cosmological data of various sorts. These early detailed studies already anticipated significant signs of dynamical DE at $3\sigma-4\sigma$ c.l. \citep{Sola:2015wwa,Sola:2016jky,Sola:2017znb,SolaPeracaula:2016qlq,SolaPeracaula:2017esw,Gomez-Valent:2018nib}. The current work aligns well with these studies,  which were like a harbinger of the dynamical character of the DE emerging from the analysis of a large body of observational data. But at the same time it leads to further insight into possibly new qualitative properties of the DE, in particular its preference for undergoing a sign flip at a nearby transition redshift, thus adopting a sort of chameleonic nature around that transition point, for the DE appears phantom-like first (viz. ``phantom matter'')  and subsequently it shows up as effective quintessence near our time (as recently observed by DESI). This combined feature of the DE optimizes to a large extent the quality fit to the overall cosmological data as compared to the $\CC$CDM. Now while this positive feature holds true for the two sorts of BAO, it is not so for curing the $H_0$ tension (see below).

All in all, model $w$XCDM potentially offers a rich conceptual framework for our understanding of the physical phenomena that may be involved.  In the $w$XCDM, the formation of PM bubbles is induced by quantum fluctuations associated to the nearing of the universe towards a de Sitter phase. The phenomenon need not be unique: the bubbles of PM  could well be operating more than once at earlier times on the same physical grounds, what would trigger an anomalous outgrowth of structures at even higher redshifts, say in the range  $z\sim 5-10$. This might explain the appearance of the large scale structures recently spotted at unusually high redshifts by the JWST mission\,\citep{Labbe:2022ahb,Adil:2023ara,Menci:2024rbq}.  Such LSS `anomalies', which find no explanation in the $\CC$CDM, might also be described within the current proposal since they can be conceived as being the earliest bubbles of PM popping up in the late universe, corresponding probably to the first tunneling attempts anticipating the eventual dS phase near our time. This is reasonable, given the fact  that the tunneling process towards the final de Sitter era should be gradual, exactly as in the early universe \citep{Mavromatos:2021urx}.

Last but not least, as promised, we would like to advance here, if only very briefly, the outcome of the preliminary analysis performed with BAO 3D data. Our extended study has revealed that the $w$XCDM model with BAO 3D (replacing BAO 2D) data improves once more (and quite significantly)  the overall fit to the cosmological data as compared to the $\Lambda$CDM, see\,\citep{LXCDM2024}. This should not come as a big surprise, if we take into account our discussion in the previous section, where we have shown by appealing to  Table \ref{tab:table_chi2} in the Appendix that the composite DE models beat the $\Lambda$CDM in every single observable, not only in those driving the tensions. It means that the composite feature of the DE proves really fruitful to better accommodate the various sources of cosmological data. This is remarkable in the first place,  for it means that the superiority of the composite DE models over the $\CC$CDM is not compromised at present by the particular type (2D  or 3D) of BAO  data used, despite them being currently under some tension \citep{Camarena:2019rmj,Gomez-Valent:2023uof,Favale:2024sdq}. Notwithstanding this relevant fact, which holds good for the global  fit involving all sorts of cosmological data, we should not underemphasize that, insofar as concerns the $H_0$ tension itself, the fit with  BAO 3D data proves unable to match the efficiency of the BAO 2D fit. Specifically, with BAO 3D the $H_0$ tension persists at a level of $\sim 2.9\sigma$ (no longer at the insignificant level of $0.25\sigma$ like in the BAO 2D case).  Thus, while it is true that with BAO 3D the $H_0$ tension gets somewhat reduced from the large initial value ($\sim 5\sigma$ c.l.),  it remains still sizeable.  In contrast, we find that the growth tension is less affected.  All in all, a final solution to the cosmological tensions -- above all to the $H_0$ tension --  is still being compromised by a lack of consistency between the two sources of BAO, angular and anisotropic.  The systematic exposition of the results with BAO 3D, along with the quantitative differences with respect to the analysis using BAO 2D, is beyond the scope of this work and its presentation is left for a separate publication\,\citep{LXCDM2024}.

At the end of the day,  it is encouraging to realize that the physics involved in the PM scenario presented here for a potential solution to the cosmological tensions, might ultimately stem from (dynamical) quantum vacuum phenomena, what gives hope  for eventually achieving a deeper understanding of the cosmological evolution of the universe from  fundamental physics. Whether this is actually the case or not will depend,  of course,  on the eventual resolution of the BAO tension itself, which appears to be the main stumbling block for the ultimate understanding of the problem.

\section{Acknowledgements}
This work is  partially supported by grants PID2022-136224NB-C21 and  PID2019-105614GB-C21, from MCIN/AEI/10.13039/501100011033.  AGV is funded by “la Caixa” Foundation (ID 100010434) and the European Union's Horizon 2020 research and innovation programme under the Marie Sklodowska-Curie grant agreement No 847648, with fellowship code LCF/BQ/PI21/11830027.  JSP is funded also  by  2021-SGR-249 (Generalitat de Catalunya) and
CEX2019-000918-M (ICCUB, Barcelona).  Both of us  acknowledge networking support by the COST Association Action CA21136 ``{\it Addressing observational tensions in cosmology
with systematics and fundamental physics (CosmoVerse)}''.

\appendix
\section*{Full triangle plot and breakdown of $\chi^2_{\rm min}$ contributions}

\begin{figure}[ht!]
    \centering
    \includegraphics[scale=0.12]{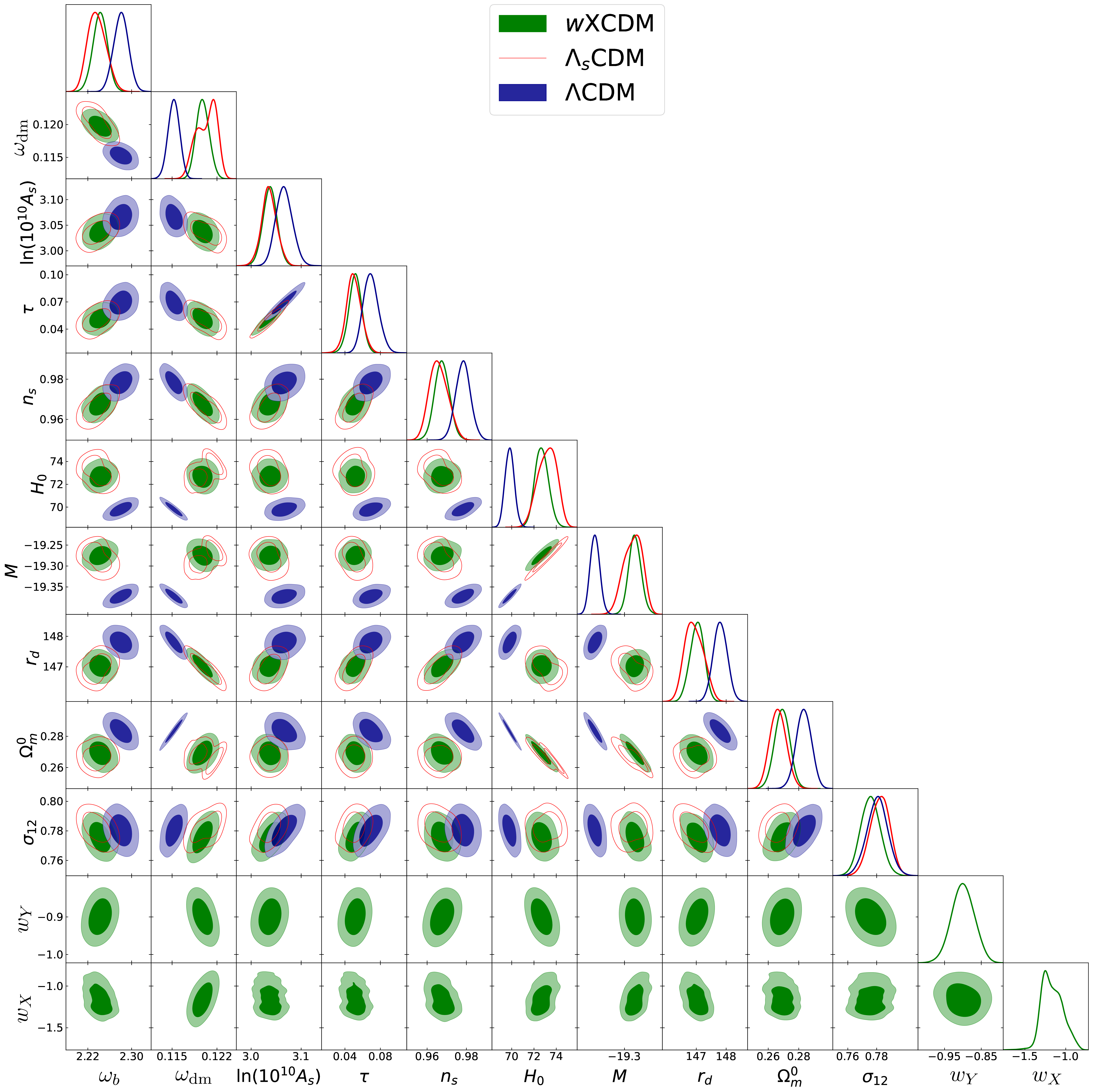}
    \caption{\scriptsize Full triangle plot for the various models studied in this paper. We show the constraints at 68$\%$ and $95\%$ CL in all the relevant planes of the parameter spaces, together with the individual one-dimensional posterior distributions. $H_0$ is given in km/s/Mpc.}
    \label{fig:full_triangle_plot}
\end{figure}

\begin{table}[ht!]
\centering
\renewcommand{\arraystretch}{2}
\begin{tabular}{|c ||c |c |c |}
\hline
{\small $\chi^2_i$} & {\small $\Lambda$CDM}   & {\small $w$XCDM }  & {\small $\Lambda_s$CDM}
\\\hline
$\chi^2_{\rm Planck\_highl\_TTTEEE}$ & 2365.09  & 2350.78 & 2350.81 \\\hline
$\chi^2_{\rm Planck\_lowl\_EE}$ & 404.20  & 395.91 & 396.07 \\\hline
$\chi^2_{\rm Planck\_lowl\_TT}$ & 21.34   & 25.21 & 27.80 \\\hline
$\chi^2_{\rm Planck\_lens}$ & 10.77   & 9.31 & 10.74 \\\hline\hline
$\chi^2_{\rm CMB}$ & 2801.40 & 2781.21 &  2785.42 \\\hline
$\chi^2_{{\rm Pantheon+SH0ES}}$ & 1312.83  & 1290.11 & 1302.21 \\\hline
$\chi^2_{\rm BAO}$ & 23.87 &   13.27 & 13.80 \\\hline
$\chi^2_{f\sigma_{12}}$ & 15.99  & 12.68 & 9.41 \\\hline
$\chi^2_{\rm CCH}$ & 12.67 & 10.52 & 10.20\\\hline\hline
$\chi^2_{\rm min}$ & 4166.76 & $4107.62$ & $4120.04$ \\\hline
\end{tabular}
\caption{\scriptsize Individual $\chi^2_i$ contributing to $\chi^2_{\rm min}$, obtained in the fitting analyses  for the various models with CMB+CCH+SNIa+SH0ES+BAO+$f\sigma_{12}$. $\chi^2_{\rm CMB}$ contains the total CMB contribution, i.e. it is the sum of all the Planck $\chi^2_i$, which we list in the upper half of the table.}
\label{tab:table_chi2}
\end{table}

\newpage
\newpage

\bibliographystyle{aasjournal}

\end{document}